\documentclass[reprint, amsmath,amssymb,aps,letter]{revtex4-1}

\usepackage{float}
\usepackage{graphicx}
\usepackage{dcolumn}
\usepackage{bm}

\usepackage[usenames,dvipsnames]{color}

\usepackage{longtable}
\usepackage{multirow}
\usepackage{comment}
\usepackage{soul}
\usepackage[normalem]{ulem}

\begin{document}

\preprint{APS/123-QED}


\title{Brillouin light scattering studies of aqueous {\it{E. coli}} cell lysate:  Viscoelastic properties of a multimacromolecular solution}

\author{D. F. Hanlon}
\email{dfh031@mun.ca}

\author{G. Todd Andrews}
\affiliation{Department of Physics and Physical Oceanography, Memorial University, St. John's, NL, Canada, A1B 3X7}

\author{S. Heravi and V. Booth}
\affiliation{Department of Biochemistry, Memorial University, St. John's, NL, Canada, A1C 5S7}

\date{\today}
\begin{abstract}
Brillouin spectroscopy was used to probe the viscoelastic properties of \textit{E. coli} bacterial cell lysate in aqueous solution at GHz-frequencies over the range -5.0 $^\circ$C $\leq T \leq$ 50.0 $^\circ$C. This work offers a first temperature dependent study on cell lysate by Brillouin light scattering. A single peak was observed in the spectra and attributed to a longitudinal acoustic mode of the solution. The speed of sound, bulk modulus, apparent viscosity and hypersound attenuation were extracted from the frequency shift and FWHM of the spectral peak. This study demonstrate that the behavior of complex multimacromolecular solutions, as shown by \textit{E. coli} lysate, can exhibit viscoelastic properties closely akin to those observed in simple binary aqueous protein solutions. Furthermore, our findings show that by analyzing the raw spectral signature of the Brillouin spectra, it may be possible to identify protein denaturation.

 
\end{abstract}

\pacs{Valid PACS appear here}
\maketitle

\section{Introduction} 
Cell lysate is a complex mixture of biomolecules that is released when cells are ruptured. It finds wide application in various fields of research, including fundamental biological studies and biotechnological applications like protein purification \cite{chiang2005application}.  Bacterial cell lysate, in particular, has been extensively investigated in cell-free metabolic engineering for producing target molecules and debugging pathways, offering potential advantages for engineering bacterial cells such as the commonly used lysate derived from \textit{Escherichia coli} (\textit{E. coli}) \cite{kay2015lysate}. The lysate derived from \textit{E. coli} has been used extensively in the past due to its well-characterized genetics, ease of cultivation, and relevance to both environmental and clinical settings of the parent bacterium \cite{macnab1996escherichia}. The study of \textit{E. coli} lysate has yielded valuable insights into the composition, molecular interactions, and biochemical processes within bacterial cells, thanks to the diverse assortment of biomolecules it contains \cite{molloy2000proteomic}. These biomolecules, including proteins, nucleic acids, lipids, carbohydrates, and associated metabolites, play crucial roles in essential cellular functions like metabolism, gene expression, signal transduction, and cell division \cite{molloy2000proteomic}.
 
In the context of experimental fluids physics, cell lysate serves as a model solute for studying aqueous multimacromolecular solutions commonly found in biological systems. Understanding these solutions has proven challenging due to their complex composition, however, their complexity also makes the system interesting as it consists of a diverse array of molecules differing in size, charge, and shape. To investigate the diffusivity of proteins and polymer crowding agents in cell lysate solutions, nuclear magnetic resonance techniques have been employed \cite{trosel2023diffusion, wang2010effects}. In one study, the rotational and translational diffusion of a 7.4 kDa test protein, chymotrypsin inhibitor 2, was examined in the presence of synthetic polymer crowders, protein crowders, and lysate \cite{wang2010effects}.  The results indicated that artificial crowders like poly(vinylpyrrolidone) 40 and Ficoll do not accurately replicate cellular environments due to the absence of many non-covalent interactions that are present in real cells. Another study investigated the effects of polymer crowders on polyethylene glycol in bacterial cell lysate, revealing that polyethylene glycol is also unsuitable as an excluded-volume crowding agent due to its strong association with lysate components \cite{trosel2023diffusion}. Rheology techniques have also been employed to investigate the low-frequency viscoelasticity of cell lysate solutions \cite{wang2010effects,newton2017investigating, trosel2023diffusion}. These studies focused on shear and storage moduli, as well as viscosity in relation to shear rate. Interestingly, all of these parameters exhibited a slight decrease with increasing shear rate. The high-frequency viscoelastic properties of cell lysate solutions, however, have not been examined. Such high-frequency viscoelastic characterization would inform the testing and refinement of theoretical models based on simpler systems to ensure that they capture the essential physics necessary to describe the behavior of more complex systems. \\
\indent
This paper reports the results of Brillouin light scattering experiments with an aqueous solution of \textit{E. coli} cell lysate over the temperature range -5.0 $^\circ$C $\leq T \leq$ 50 $^\circ$C. 
This is the first study to investigate the GHz-frequency viscoelastic properties of a cell lysate solution.  We determined the hypersound velocity, bulk modulus, apparent viscosity and hypersound attenuation as a function of temperature. The results of this work provide new insight into the physics of complex multimacromolecular solutions that mimic real-world biological systems.

\section{Methodology}
\subsection{Solution Preparation} 


A 500 mL flask containing 75 mL of Lysogeny Broth (LB) media, consisting of 5 g of yeast extract, 10 g of NaCl, and 10 g of tryptone \cite{fisher}, was closed with a sponge plug and aluminum foil, and autoclaved for sterilization. After reaching room temperature, overnight cultures of \textit{E. coli} JM109 were prepared by inoculating the 75 mL of LB media with 1 mL aliquots of frozen glycerol cell stocks. The inoculated cultures were placed in a 30 $^\circ$C incubator with shaking at 150 rpm, protected with a foam stopper. Six 4 L flasks, each containing 1 L of LB media, were pre-warmed to 37 $^\circ$C. The following day, the flasks were inoculated with 1 mL of overnight culture per 100 mL of fresh LB media and incubated at 37 $^\circ$C with shaking at 175 rpm. Each culture was protected with a foam stopper.

Cells were harvested during the midlog phase at an absorption of 600 nm (A$_{600}$) ranging from 0.6 to 1.0. The absorption is a measure of the optical density (OD) and is often used to characterize the concentration of cells in a culture. The higher the OD implies a higher cell concentration \cite{widdel2007theory}. The harvested cells were centrifuged at 5670 g for 20 minutes at 4 $^\circ$C. The resulting pellet was subjected to three cycles of the French press at a pressure of 10,000 psi (6.9 $\times$ 10$^7$ Pa) and maintained at 4 $^\circ$C to mechanically lyse the cells. This method of lysing the cell involves passing the pellet through a tiny valve at extremely high pressures. As the cells leave the valve, the pressure difference between the valve pressure and the surrounding atmospheric pressure generates significant shear stress, which eventually causes the cells to burst. The volume of cell lysate obtained was approximately 30 mL. Once complete, the lysate was subject to lyophilization to remove water. To begin this process, the lysate was stored in a freezer set at -80$^\circ$C for 24 hours. Subsequently, the lysate is transferred into a vacuum chamber, where it undergoes freeze-drying over the duration of 48 hours. Once freeze-dried, the volume reduced to $\sim5$ mL. The lyophilized cell lysate was stored in a 15 mL falcon tube at -20 $^\circ$C for future use.

To prepare the cell lysate solution, approximately 0.3 g of cell lysate was carefully added to a glass sample cell. The lysate was dissolved in  4.0 $\pm$ 0.1 ml of water. This gives a water concentration comparable to that inside cells. The solution was then thoroughly mixed for approximately 10 minutes using a stirring stick to promote uniform dispersion of the components. Following the completion of the mixing step, Brillouin scattering experiments were promptly initiated.

The density of the aqueous cell lysate solution was determined to be $\rho = 1.0 \pm 0.1$ g/cm$^3$, using the equation for a two-component solution
\begin{equation}
    \rho = \frac{m_W + m_L}{ \frac{m_W}{\rho_w} + \frac{m_L}{\rho_L} },
    \label{eq:density}
\end{equation}
where $\rho_W = 0.997$ g/cm$^3$ is the density of water, $\rho_L = 1.41$ g/cm$^3$ (dried) \cite{fischer2004average} is the density of cell lysate, and $m_W$ and $m_L$ are the mass of water and cell lysate (dried), respectively.


\subsection{Brillouin Spectroscopy} \label{BLS}
\begin{figure*}[t]
    \centering \includegraphics[scale = 0.45]{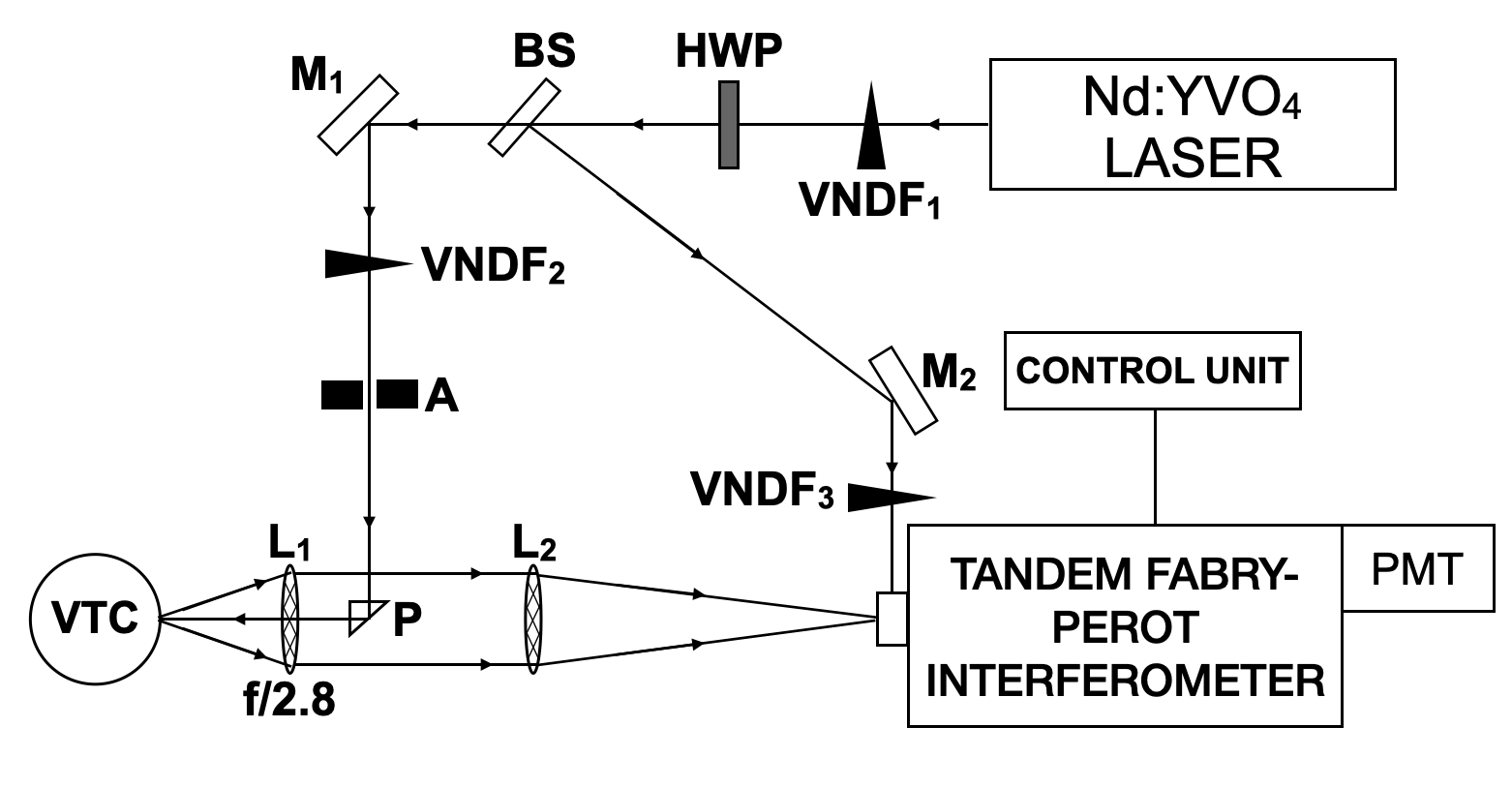}
    \caption{Optical setup for Brillouin scattering used in this study. VNDF - Variable neutral density filter, HWP - half wave plate, BS - beam splitter, M - mirror, VNDF - variable neutral density filter, A - aperture L - lens, P - prism, VTC - variable temperature chamber, PMT - photomultiplier tube.}
    \label{fig:BrillouinSetup}
\end{figure*}
Brillouin spectroscopy is a technique based on inelastic light scattering that enables the investigation of thermal acoustic phonons within a medium. For a $180^{\circ}$ backscattering geometry as used in the present work, application of energy and momentum conservation to the scattering process reveals that the phonon velocity, $v$, and frequency shift of the Brillouin peak, $f_B$, are related by  
\begin{equation}
v = \frac{f_B\lambda_i}{2n},
\label{eq:brillouineqn}
\end{equation}
\noindent
where $n$ is the refractive index of the target material at the incident light wavelength $\lambda_{i}$. The refractive index of the cell lysate solution used in the present work was taken to be equal to that of water, $n$ = 1.33.

The apparent viscosity $\eta = 4\eta_{s}/3 + \eta_{b}$, where $\eta_{s}$ and $\eta_{b}$ are the shear and bulk viscosity, respectively, is related to the full width at half maximum (FWHM) of the Brillouin peak due to the longitudinal acoustic mode via \cite{rouc1976},
\begin{equation}
    \Gamma_B = \frac{16\pi^2 n^2}{\rho \lambda_i^2} \left[\eta + \frac{\kappa}{C_p}(\gamma - 1) \right], 
    \label{eq:FWHM}
\end{equation}
where $\rho$ is the density, $\kappa$ is the thermal conductivity, and $\gamma = C_p/C_v$ is the ratio of specific heat at constant pressure to that at constant volume.  The second term in the brackets of this expression is usually neglected for simple liquids, thus leaving the expression for viscosity \cite{rouc1976},
 \begin{equation}
\eta = \frac{\rho \lambda_i^2}{16 \pi^2 n^2}\Gamma_B.
    \label{eq:viscosity}
\end{equation}

Knowledge of the hypersound velocity and apparent viscosity permits the real and imaginary parts of the complex longitudinal modulus, $M = M^{\prime} + iM^{\prime\prime}$, to be determined through the relations

\begin{equation}
    M^{\prime} = \rho v^2
    \label{eq:storagemod}
\end{equation}
and
\begin{equation}
    M^{\prime\prime} = 2 \pi \eta f_B,
    \label{eq:lossmod}
\end{equation}
where $M^{\prime}$ and $M^{\prime\prime}$ are referred to as the storage and loss moduli, respectively \cite{bailey2019brillouin}. 

The frequency-independent sound absorption coefficient may also be determined from Brillouin data through \cite{rouc1976},
\begin{equation}
\frac{\alpha}{f^2} = \frac{\Gamma_B}{2vf_B^2}.
\label{eq:absorp}
\end{equation}

Brillouin spectra were collected using the $180^\circ$ backscattering setup depicted in Fig. \ref{fig:BrillouinSetup}. A single-mode Nd:YVO$_4$ solid-state laser (Coherent Verdi-2) with an output power of 1.66 W and an emission wavelength of 532 nm served as the incident light source. A half-wave plate (HWP) was used to rotate the plane of polarization from vertical to horizontal.  After leaving the half-wave plate, the beam was divided into a reflected reference beam and a transmitted sample probe beam by a beam splitter (BS). The reference beam is used in conjunction with a mechanical shutter system to maintain interferometer alignment and to prevent saturation of the photomultiplier tube (PMT) by intense elastically scattered light from the sample.  The transmitted beam was reflected through an angle of 90$^\circ$ by front surface mirror (M1) and incident on a right-angle prism (P) where it underwent total internal reflection. The probe beam was then focused onto the sample using a camera lens (L1) with a focal length of 5 cm and an f/\# of 2.8. Two variable neutral density filters (VNDF1 and VNDF2) were placed in the beam path to reduce the power at the sample to $\sim100$ mW.  Scattered light was collected and collimated by the same camera lens and subsequently focused onto the 450 $\mu$m-diameter input pinhole of a six-pass tandem Fabry-Perot interferometer by a $f=40$ cm lens (L2).  The frequency-analyzed scattered light transmitted by the interferometer was incident on the photocathode of a photomultiplier tube (PMT) where it was converted to an electrical signal and sent to a computer for recording and display. 

The sample was housed in a custom-built variable temperature chamber (VTC). A Lakeshore Cryotronics temperature controller was used in conjunction with a resistor and thermoelectric cooler to regulate sample temperature. Depending on the target temperature for the experiment, either the heater or the cooler was initially activated. Once the desired temperature was reached, the heater and cooler worked in tandem to maintain the sample at that temperature. If the temperature exceeded the desired setpoint, the heater was shut-off and the cooler was switched on to reduce the temperature to the setpoint. Conversely, when the sample dropped below the setpoint, the cooler was turned off and the heater was activated to return the temperature to the setpoint. The accuracy of this variable temperature chamber was $\pm 0.4$ $^\circ$C.  Further details can be found in Ref.~\cite{hanlon2023temperature}.

\section{Results and Discussion}
\subsection{Brillouin Spectrum}
\subsubsection{General Features}
\begin{figure}[t]
    \centering \includegraphics[scale=0.38]{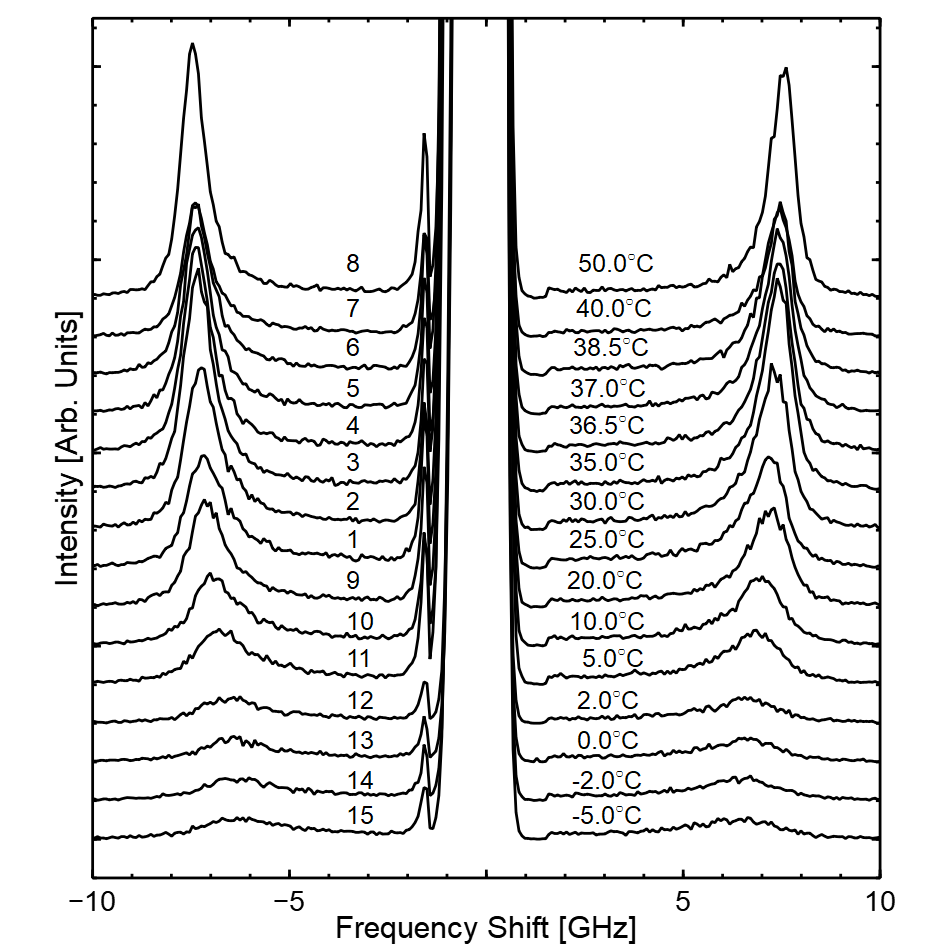}
    \caption{Brillouin spectra of an {\it{E. coli}} bacterial cell lysate-water solution collected at the temperatures indicated. Integer numbers (1-15) represent the sequential order in which spectra were collected. }
    \label{fig:Ecoli_Spectra}
\end{figure}

Figure \ref{fig:Ecoli_Spectra} shows Brillouin spectra collected from the \textit{E. coli} lysate solution at temperatures ranging from -5.0 $^\circ$C to 50 $^\circ$C. The chronological order in which the spectra were collected is also indicated. A single Brillouin doublet was observed in the spectra at $\sim\pm7.2$ GHz and attributed to the longitudinal acoustic mode of the cell lysate solution. Brillouin peak frequency shifts and linewidths (FWHM) were obtained by fitting Lorentzian functions to the Stokes and anti-Stokes Brillouin peaks and averaging the resulting best-fit parameters. To obtain linewidths, the instrumental linewidth of 0.3 GHz was subtracted from FWHM values obtained from the Lorentzian fits. Estimated uncertainties in peak parameters (frequency shift and FWHM) were obtained from the uncertainty in the Lorentzian fits.

\subsubsection{Sedimentation Effects} \label{sec:ExpChallenges} 
It was noted that the frequency shifts of Brillouin peaks in the first three spectra (obtained at 25 $^\circ$C, 30 $^\circ$C, and 35 $^\circ$C and collected consecutively within an hour after solution preparation) trended toward that of water, raising concerns about the integrity of the data collected. The reason for this was later discovered to be sedimentation of larger particles present in the cell lysate, as made apparent by the presence of an $\sim2$ mm-thick white residue at the bottom of the sample cell when removed from the temperature-controlled chamber at the conclusion of the experiment. These larger particles likely include remnants of the bacterial cell wall (characteristic size of $\sim1$ $\mu$m \cite{riley1999correlates,huang2008cell}), and not molecular constituents such as proteins, nucleic acids, or lipids due to the small sizes of the latter (see Table \ref{tab:cellcomponents}).  There is also a small percentage of smaller particles such as ions, metabolites, and polysaccharides \cite{johnson2002molecular} not included in Table \ref{tab:cellcomponents} due to their minimal abundance and significant variation in radii and density.  This is in accord with Stokes' Law which predicts that the terminal velocity of a particle in a fluid is proportional to the square of its radius \cite{batchelor1967introduction} (the strongest dependence on any parameter in this equation).  With this in mind, the sedimentation time was estimated by first stirring the solution for approximately 10 minutes until homogeneous (as for the as-prepared solution at the outset of the experiment) and then visually monitoring it until the larger particles had settled. This time was found to be $\sim45$ minutes. It is therefore likely that the spectra collected at 25$^\circ$C, 30$^\circ$C, and 35$^\circ$C were affected by sedimentation. This was considered when analyzing the experimental results. It should also be noted that the probing incident laser beam was not impeded by the settled residue.

\begin{table}[t]
  \centering
  \caption{The approximate radius and density of cellular components in bacterial cells, along with their composition by weight \cite{johnson2002molecular}.}
  \begin{tabular}{lcccr}
    \hline \hline \vspace{0.5mm}
    Molecule & Radius & Mass Density  & Abundance\\
     &  (nm) & (g/cm$^3$) & \% Dry weight \\
    \hline
    Proteins &  2 - 10 \cite{zhdanov2009conditions} & $\sim$ 1.35 \cite{fischer2004average} & 50 - 55 \cite{feijo2013intracellular}\\
    RNA & 7 - 10 \cite{borodavka2016sizes} & $\sim$ 2.00 \cite{anderson1966separation}&  20 \cite{feijo2013intracellular} \\
    Phospholipids & 0.5 - 1 \cite{nagle2000structure} & $\sim$ 1.01 \cite{johnson1973osmotic} & 7 - 9 \cite{feijo2013intracellular}\\
    DNA &3 - 5 \cite{sinden1994dna} & $\sim$ 1.70 \cite{anderson1966separation,panijpan1977buoyant} & 3 \cite{feijo2013intracellular} \\
    \hline \hline
  \end{tabular}
  \label{tab:cellcomponents}
\end{table}

\subsubsection{Temperature Dependence}

\begin{figure}[t]
    \centering
     \includegraphics[scale=0.42]{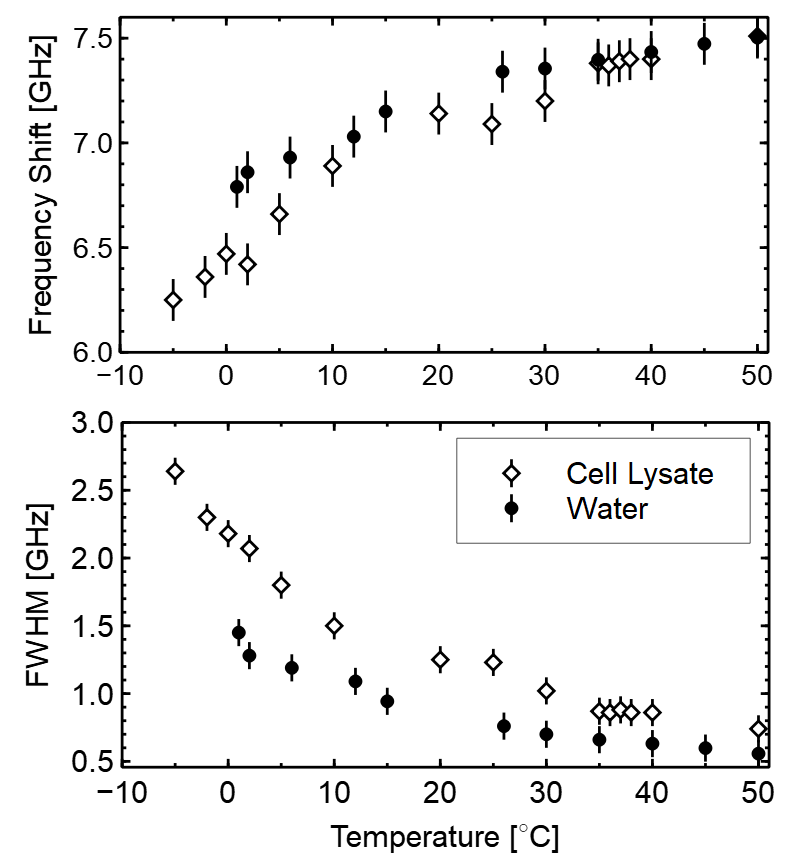}
    \caption{Plot of frequency shift and FWHM as a function of temperature for water and an aqueous {\it{E. coli}} cell lysate solution. Error bars represent the uncertainty in the spectral parameters.}
    \label{fig:Freq_FWHM_CellLys}
\end{figure}

Figure \ref{fig:Freq_FWHM_CellLys} shows the temperature dependence of the Brillouin peak frequency shift for water and the cell lysate solution. In general, the frequency shift for the lysate solution increases with increasing temperature. Previous Brillouin scattering work on aqueous protein systems have consistently reported a similar trend \cite{bailey2019brillouin,comez2012,comez2016,lupi2011,palombo2019brillouin,monaco2001glass,hanlon2023temperature}. Notably, at higher temperatures, the frequency shift for the lysate solution has values that closely resemble those of water.  For temperatures below $T\sim10$ $^\circ$C, however, a noticeable difference develops between the peak frequency shifts for water and the cell lysate solution. 

Figure \ref{fig:Freq_FWHM_CellLys} also shows the Brillouin peak linewidth as a function of temperature for water and cell lysate. Unlike the peak frequency shift, the FWHM for the lysate solution is different from that of water over the entire temperature range studied. This is a clear indication that even though larger particles in the cell lysate have settled-out, smaller particles such as protein and lipid molecules remain in solution. The FWHM begins to become appreciably different from that of water for $T \leq 10$ $^\circ$C, the same temperature range as noted above for the Brillouin peak frequency shift.

\subsection{Viscoelastic Properties}
\subsubsection{Temperature Dependence}
\begin{figure}[!h]
    \centering
     \includegraphics[height = 20.5cm, width=\linewidth]{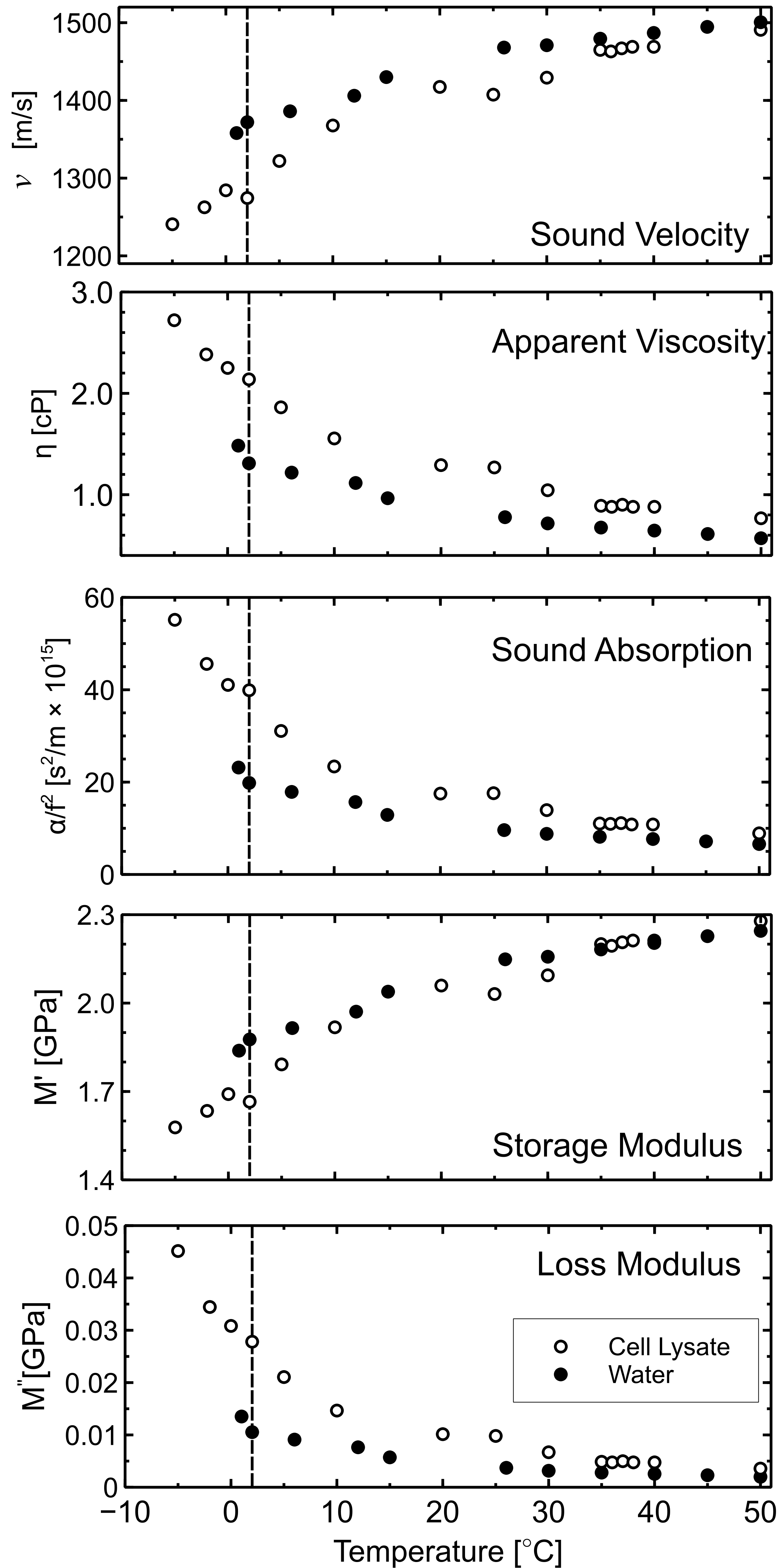}
    \caption{Sound velocity, apparent viscosity, frequency-independent sound absorption coefficient, storage, and loss modulus as a function of temperature for an aqueous cell lysate solution. The uncertainty is approximately twice the size of the symbols.}
    \label{fig:Lysate_Visco}
\end{figure}

Figure \ref{fig:Lysate_Visco} presents viscoelastic properties for water and cell lysate solution over the range -5.0 $^\circ$C to 50.0 $^\circ$C determined from the Brillouin data using equations given in Sec.~\ref{BLS}.  As can be seen, the viscoelastic properties of the cell lysate solution obtained from the first few post-sedimentation spectra (\#4 - \#8 in Fig.~\ref{fig:Ecoli_Spectra}, collected at $T>35$ $^\circ$C), are nearly independent of temperature and closely resemble those of water.  In contrast, solution viscoelastic properties obtained from spectra collected after those in this ``high-temperature'' region ({\it{i.e.}}, at $T\leq20$ $^\circ$C) are appreciably different from those of water, as can be seen in Fig.~\ref{fig:Lysate_Visco} by comparison of values for the lysate solution to those for water (direct or interpolated) at a given temperature. A possible reason for this behaviour could be the irreversible denaturation of proteins in the cell lysate solution at $T\sim40$ $^\circ$C \cite{petsko2004protein}. If this is in fact the case, this result is very interesting because it means that protein denaturation and configuration (folded or unfolded) can be detected using Brillouin spectroscopy.  This hypothesis is also supported by the results of recent molecular dynamics simulations on aqueous protein systems which show some convergence of sound velocity, adiabatic bulk modulus, and shear viscosity toward values for water for $T \gtrsim 40$ $^\circ$C and a much reduced dependence of these quantities on temperature in this range compared to that at lower temperatures \cite{hanlon2023probing}.  

The temperature dependence of several of the cell lysate solution viscoelastic properties seems to display a weak anomaly at $T\sim2$ $^\circ$C, as indicated by the dashed vertical line in Fig. \ref{fig:Lysate_Visco}.  This anomaly is apparent in the velocity, sound absorption, and storage modulus data, but barely discernible in the temperature dependence of the apparent viscosity and loss modulus.  Given that a significant fraction of the cell lysate solution is water with a well-known liquid-to-solid phase transition at 0 $^\circ$C, this anomaly is not completely unexpected and may in fact be a manifestation of a corresponding liquid-solid phase transition in the lysate solution at approximately the same temperature.

It is also interesting to note the qualitative similarity of the temperature dependence of some lysate solution viscoelastic properties to those of simple aqueous protein systems obtained using molecular dynamics simulations (see Fig.~\ref{fig:Lysate_Visco} in this paper and Figs. 2, 4 and 5 in Ref. \cite{hanlon2023probing}).  Despite considerable differences in composition, the sound velocity and storage modulus for the lysate solution and nearly all of the model systems show an overall increase with increasing temperature over the range from a few $^\circ$C to $\sim40-45$ $^\circ$C and a much weaker dependence for higher temperatures, with some convergence of values toward those for water in this upper range.  In addition, the overall decrease in shear viscosity with increasing temperature for the simulated protein-water systems mimics that of the apparent viscosity of the cell lysate solution.  Interestingly, similar behaviour was found for the temperature dependence of sound velocity, storage modulus, and apparent viscosity of natural snail mucus, a system comprised primarily, but not only, of water and glycoproteins \cite{hanlon2023temperature}.  When considered together, these results suggest that the underlying physical mechanisms responsible for the bulk viscoelasticity of simple aqueous protein solutions are also at play in multimacromolecular solutions. Consequently, it may be possible to predict the viscoelastic behaviour of complex macromolecular solutions, such as found in biological systems, using rather simple models.

\subsubsection{Activation Energy}
\begin{figure}[!h]
    \centering
    \includegraphics[scale=0.42]{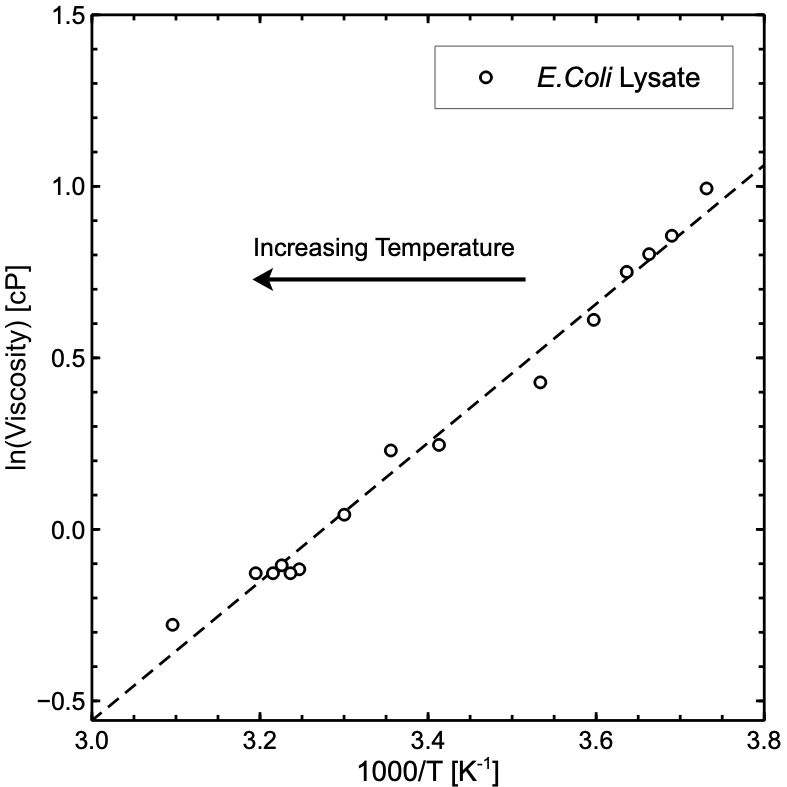}
    \caption{Natural logarithm of viscosity for an aqueous solution of bacterial cell lysate
    as a function inverse temperature. Dashed lines represent best fit. The uncertainty is approximately twice the size of the symbols.}
    \label{fig:logvisco_ecoli}
\end{figure}

Figure \ref{fig:logvisco_ecoli} shows the natural logarithm of the apparent viscosity as a function of inverse temperature. The apparent viscosity for the entire temperature region depends linearly on $1/T$ so we fit an Arrhenius relationship of the form
\begin{equation}
\eta = \eta_0 e^{E_a/k_B T} 
\label{eq:lneta}
\end{equation}
to the data. The activation energy for the cell lysate was determined to be $E_a = 17.1$ kJ/mol, which is approximately 20\% higher than the activation energy for water and other polymer-water systems \cite{lupi2011,comez2012,hanlon2023temperature}. Notably, the observed data exhibited a linear relationship, indicating a negligible thermal dependence of both the activation entropy ($\eta_0$) and enthalpy ($E_a$) across the entire temperature range studied. In contrast to other systems, no evidence of a transition to non-Arrhenius behavior was observed. In other aqueous protein solutions, the apparent viscosity typically follows an Arrhenius behavior in the high-temperature region ($T \geq 20$ $^\circ$C), which is associated with the onset of cooperative motions at the molecular level and is often described by a power law or a Vogel–Fulcher relationship \cite{lupi2011,hanlon2023temperature}. 

\section{Conclusion} 
In this study the GHz-frequency viscoelastic properties of an aqueous solution of {\it{E. coli}} cell lysate were probed by Brillouin spectroscopy across a temperature range of -5.0 $^\circ$C to 50 $^\circ$C.  We find that the observed behaviour in the frequency shift and FWHM, and consequently the speed of sound, viscosity and sound absorption closely resembles that of simple aqueous solutions containing only a single protein or polymer species. For instance, previous Brillouin scattering studies \cite{bailey2019brillouin,comez2012,comez2016,lupi2011,palombo2019brillouin,monaco2001glass,hanlon2023temperature} have shown the general increase in frequency shift and decrease in FWHM with increasing temperature. This could have potentially important implications for computer simulations and theoretical work on multimacromolecular systems because it demonstrates that the viscoelasticity of a complex systems can, at least in some cases, be accurately modelled using much simpler systems. The quantitative nature of the results obtained in this work also highlight the potential of Brillouin light scattering spectroscopy as a potent tool for examining the viscoelasticity of complex biological systems, as the influence of proteins on the viscoelastic properties of these systems is still in some ways poorly understood.

\section*{Disclosure Statement}
\noindent
No potential conflict of interest was reported by the authors.

\section*{Acknowledgments}
\noexpand
This work was partially supported by Memorial University Seed, Bridge and Multidisciplinary Fund through grant 216354-46314-2000.

\bibliography{bibliography}

\end{document}